\documentclass[twocolumn,secnumarabic,amssymb, nobibnotes, aps, prd]{revtex4-2}

\usepackage{graphicx, bm}
\begin{document}

\title{Effect of constraint and Tabu Search term on Variational Quantum Eigensolver and Subspace-Search Variational Quantum Eigensolver}%

\author{Hikaru Wakaura}%
\email{hikaruwakaura@gmail.com}
\affiliation{QuantScape Inc. 4-11-18, Manshon-Shimizudai, Meguro, Tokyo, 153-0064, Japan}

\author{Takao Tomono}%
\email{takao.tomono@ieee.org}
\affiliation{ Digital Innovation Div. Toppan Printing Co. Ltd., 1-5-1, Taito, Taito, Tokyo, 110-8560, Japan}

\date{23 March 2021}

\begin{abstract}
Subspace-Search Variational Quantum Eigensolver(SSVQE)  is searching method of multiple states  and relies on the unitarity of transformations to ensure the orthogonality of output states for multiple states.
 Therefore, this method is thought to be promising method for quantum chemistry because ordinary Variational Quantum Eigensolver (VQE) can only calculate the excited states step by step from ground state based on Variational Quantum deflation (VQD). We compare the advantage of VQE, SSVQE with/without the constraint term and/or Tabu search term, that are added by Lagrange's multiplier method so as to calculate the desired energy levels. We evaluated the advantage by calculating each energy levels of H$_2$ and HeH, respectively. As there simulation results, the accuracy calculated by constrained VQE with Tabu search indicates higher accuracy than that of our other algorithm, for analysis on H$_2$.  The accuracy calculated by constrained SSVQE indicate higher that of the constrained VQE with Tabu search. We found it is beneficial for enhance the accuracy to use constraint terms decreasing convergence times to use Tabu search terms according to the nature of molecules.  We demonstrate that constraint and Tabu search  terms contribute to the accuracy and convergence time on quantum chemical calculating.
\newline

\end{abstract}

\maketitle
\tableofcontents

\section{Introduction}\label{1} 
Quantum computer is attractive equipment because the principle is based on quantum mechanics. 
Quantum computers with a hundred qubits will be develop as quantum computer with 54 qubits that reaches quantum supremacy is announced in 2019\cite{arute_quantum_2019}.
Though it takes much time to get perfect quantum computer with more than 1-milion qubits, near-term quantum computers are used as noisy intermediate-scale quantum (NISQ) devices. The device is hybrid quantum computer that consist of classic and quantum computers, currently. Within the device, quantum information treatment is done within coherence time on quantum computer, currently. Reflecting the fact that present quantum computers are not fault tolerant, they do not have a guaranteed accuracy of the computational result\cite{Gambetta2017}.
However, such a NISQ device is believed to be considerably attractive if the gate fidelity is sufficiently high. This fact encourages us to look for practical applications of them. 

The VQE is used as one of variational quantum algorithm used in NISQ device. The algorithm is to find an approximate ground state of a given Hamiltonian H. The group read by Dr. Aran Aspuru-Guzik developed the VQE in 2011\cite{doi:10.1146/annurev-physchem-032210-103512}. This approach uses a classical optimization routine to minimize the expected energy of candidate wave functions, using the quantum device to evaluate the predicted energy\cite{PhysRevX.6.031007}. Essentially, the VQE leverages the unique capacity of quantum circuits to prepare entangled states from classical sampling. Essential ingredients of the VQE algorithm have recently been demonstrated on a variety of experimental platforms\cite{PhysRevX.8.011021}\cite{2019arXiv190210171N}\cite{2017Natur.549..242K}. These initial experiments indicate a robustness to systematic control errors (so-called coherent errors) which would preclude fully quantum calculations, as well as a possibility of scaleout of quantum circuit depth with Hamiltonian complexity. 

To expand the potential application of the VQE on the ground state, many researches have extended the method to evaluate properties of excited states of a target Hamiltonian. Several of algorithms among such extensions are the subspace-search VQE (SSVQE)\cite{PhysRevResearch.1.033062}, the multi-state contracted VQE (MCVQE)\cite{2019PhRvL.122w0401P}, and the variational quantum deflation (VQD)\cite{2018arXiv180508138H}.  The SSVQE and the MCVQE can readily evaluate the transition amplitude\cite{2020arXiv200211724I}. Besides, other algorithms to calculate the energy of molecules have been proposed\cite{2019NatCo..10.3007G}\cite{mcardle_variational_2019}\cite{PhysRevLett.122.140504}. On the other hands, we have limitation of application of VQD method since the transition amplitude is related to properties of the system such as the absorption or emission spectrum of photon.
The SSVQE is the method that takes two or more orthogonal states as inputs to a parametrized quantum circuit, and minimizes the expectation value of the energy in the space spanned by those states.  This method automatically generates the orthogonal condition on the output states, and is possible for us to remove the swap test\cite{2013PhRvA..87e2330G}, which has been employed in the previous works to ensure the orthogonality.

Constrained Algorithm is introduced to Loop Quantum Gravity system as Master constrained algorithm \cite{2006PhLB..635..225H}. This constrained condition is developed for overcoming the complications associated with Hamiltonian constrain of the Dirac algebra. We can control the complications by moving of spin pair in quantum algorithm.
These are three 'no-go's' that are all well-known consequences of standard nonrelativistic Hilbert space quantum theory. However, like Einstein's radical re-derivation of Lorentz's transformation based upon privileging a few simple principles, we here introduce the above constraints term to the level of fundamental information-theoretic 'laws of nature' from which quantum theory can, we claim, be deduced. 

The tabu search (TS) algorithm\cite{Tabu} are one of the meta-heuristic search algorithms devised by Fred Glover. Therefore, it is simply implemented method to iteratively find a near-optimal solution, and it has been successfully used to solve various optimization problems. For example, optimization problem of network used for telecommunication can be solved by TS with Quantum Approximate Optimization Algorithm(QAOA) efficiently\cite{2020arXiv201109508M}. The TS algorithm with a flexible memory system has the ability to climb out of local minima, suffers from the tuning of  the tabu tenure, meaning that it still becomes stuck at local minima and has a low speed of convergence.
It takes a great deal of time to escape to near-global optimal from current position. Therefore, both intensification and diversification strategies should be considered to improve the robustness, effectiveness and efficiency of simple TS; a more powerful neighborhood structure can be feasibly constructed by applying quantum computing concepts.

In this paper, we compared to effect of complex complex complex algorithm that are VQE, SSVQE with/without constraint term and/or tabu search term for molecules of H$_2$ and HeH. Simulation methods are introduced firstly. Next, initialization will be explained in detail. After these simulation, we denote the simulation results and discuss about utilization of these methods based on the results. We conclude that constraint and Tabu search terms are beneficial. 

\section{Methods}\label{2} 
In this section we describe VQE and SSVQE with/without constraint term and /or tabu search term. Firstly, we explain how to find the minimum value of excited energy levels by VQE method and its flowchart with quantum circuit to perform it including equation. Secondary, we explain about SSVQE method. 
\subsection{Variational Quantum Eigensolver (VQE) method}\label{2-1} 
A flowchart outline of the VQE algorithm is shown in Fig. 1. We must prepare initial parameter set ${\bm \theta}_{i=0}$ on classical computer and then we calculate eigenvalue $E({\bm \theta}_{i})$ of energy by using ansatz on quantum circuit, and then the classical computer decide whether or not the eigenvalue $E({\bm \theta}_{i+1})$ is smaller than settled minimum eigenvalue $E_0$. If the eigenvalue $E({\bm \theta}_{i+1})$ is not smaller than minimum eigenvalue $E_0$, optimization of parameter set ${\bm \theta}$ is done by using Powell method on classical computer. After optimization, new eigenvalue $E({\bm \theta}_{i+1})$ is substituted for ansatz in the quantum circuit instead of $E({\bm \theta}_{i})$.
When a determination condition $(E({\bm \theta}_{i+1})=E_0)$ is not satisfied, the above processing is repeated until the determination condition is satisfied.
An quantum algorithm is presented by second-order quantum hamiltonian $H$ as follows.
\begin{equation}
H = \sum_{j,k = 0}h_{jk}c_j^\dagger c_k + \sum_{j,k,l,m = 0}\langle jk \mid\mid lm \rangle c_j^\dagger c_k^\dagger c_l c_m \label{molham}.
\end{equation}

To calculate the matrix elements $E$, we use the quantum circuit to evaluate the inner products,
\begin{equation}
E=\langle \Phi \mid H \mid \Phi \rangle\label{vqearchitype},
\end{equation}

where $\Phi$ is taken to be the initial approximate ground state $\Phi$. Optimization is performed by solving ground state $\Phi$ to minimize $E_0$. The state is represented by Slater's determinant represented by occupied and/or unoccupied orbitals.
For example, we consider the case of 4-qubits system. Each one qubit is represent as $\mid 0\rangle$ and $\mid 1\rangle$. One state of the ground slater determinant is represented as $\mid1100\rangle$ because there are two occupied bonding orbitals and two unoccupied anti-bonding orbitals. 
Here, 1 indicates the orbital is occupied and 0 indicates the orbital is unoccupied, respectively. An left hand side 11 of $\mid1100\rangle$ is bonding orbitals and right hand side of that is anti-bonding orbitals. Each term of hamiltonian eq.1 is one-body integral and two-body integral, respectively. Indices of them indicate the index of the orbital. In general, any materials are stable when their energy level is lowest. Comparing the energy of bonding orbital to anti-bonding orbital, bonding orbital is more stable. hydrogen molecules are in the ground state when their two electrons are both in the bonding orbitals as up-spin and down-spin pair. Bonding orbitals are referred to $\sigma$ bonding orbitals. The bonding orbital is expressed for the basis function of STO-3G. All terms of UCC and hamiltonian are expressed as Pauli operator by Jordan-Wigner or Bravyi-Kitaev transformation\cite{doi:10.1021/acs.jctc.8b00450,2017arXiv171007629M}. 
A transformed hamiltonian is represented by
\begin{eqnarray}
H &=&f_01+f_1\sigma^z_0 +f_2\sigma^z_1 +f_3\sigma^z_2 +f_1\sigma^z_0\sigma^z_1 \\ \nonumber
&+& f_4\sigma^z_0\sigma^z_2 +f_5\sigma^z_1\sigma^z_3 \\ \nonumber
&+&f_6\sigma^x_0\sigma^z_1\sigma^x_2 +f_6\sigma^y_0\sigma^z_1\sigma^y_2\\ \nonumber
&+& f_7\sigma^z_0\sigma^z_1\sigma^z_2
+f_4\sigma^z_0\sigma^z_2\sigma^z_3 +f_3\sigma^z_1\sigma^z_2\sigma^z_3 \\ \nonumber
&+& f_6\sigma^x_0\sigma^z_1\sigma^x_2\sigma^z_3 +f_6\sigma^y_0\sigma^z_1\sigma^y_2\sigma^z_3 +f_7\sigma^z_0\sigma^z_1\sigma^z_2\sigma^z_3
\label{bkham}.
\end{eqnarray}

Generally speaking, we compute excited-states energy after calculating ground-state energy on quantum chemistry calculation.
We prepare cluster terms T to carry out transition from a ground state to an excited state as we can not calculate wave function on excited-state by only hamiltonian. The cluster is called as Unitary Coupled Cluster (UCC). We apply Unitary Coupled Cluster of Single and Double (UCCSD) on the condition of single and double excitation terms\cite{2018PhRvA..98b2322B}.
Cluster terms T is expressed as, 

\begin{equation}
T=\sum_{j\in occu.,k\in vac.}\theta_{k}^{j}c_j^\dagger c_k + \sum_{j,k\in occu.,l,m\in vac.}\theta_{lm}^{kj}c_j^\dagger c_k^\dagger c_l c_m
\label{clus}.
\end{equation}

We introduce ansatz on quantum circuit to multiply Hamiltonian by $exp(i(T - T^{\dagger}))$ for excited state energy. After that, Hamiltonian and Cluster terms are decomposed by Suzuki-Trotter transformation \cite{McClean_2016}. As the depth of circuit (repeating ratio) increases, the variable coefficients become smaller. 
If the depth of circuit is much larger than 100, we will not need variable coefficient. In the case, calculation time become infinite. Therefore, the depth had better be set two. As preparation of optimization, we seek $\theta_k$ of Pauli operator by using $exp(-i\theta_kP_jt)$ on quantum  computer.  $\theta_k$ is variable coefficient of k-th term, $P_j$ is j-th Pauli operator, and t is the coefficient. We can prepare to make excited state energy.

\begin{figure*}
\includegraphics[scale=0.7]{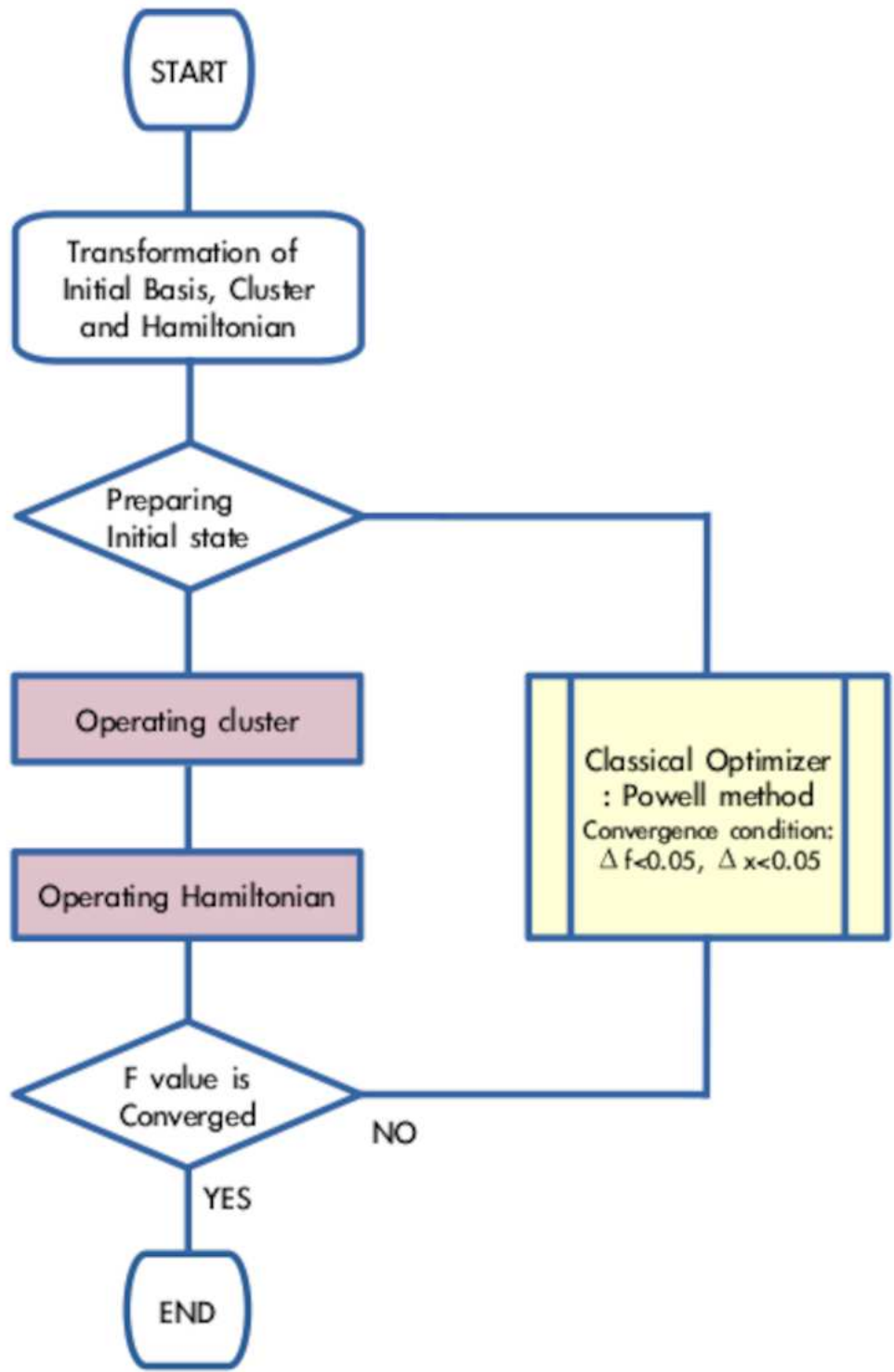}
\newline
\newline
\caption{Flowchart of VQE method. Purple area indicates the processes performed by quantum computer and yellow area indicates the processes performed by classical computer, respectively.
Quantum computer calculate the energy levels, constraint terms, Tabu search, and deflation terms. Classical computer optimizes the parameter set $\bm{\theta}$ by Powell method\cite{Powell1964}.
}
\label{vqediag}
\end{figure*}

\begin{figure}[h]
\includegraphics[scale=0.7]{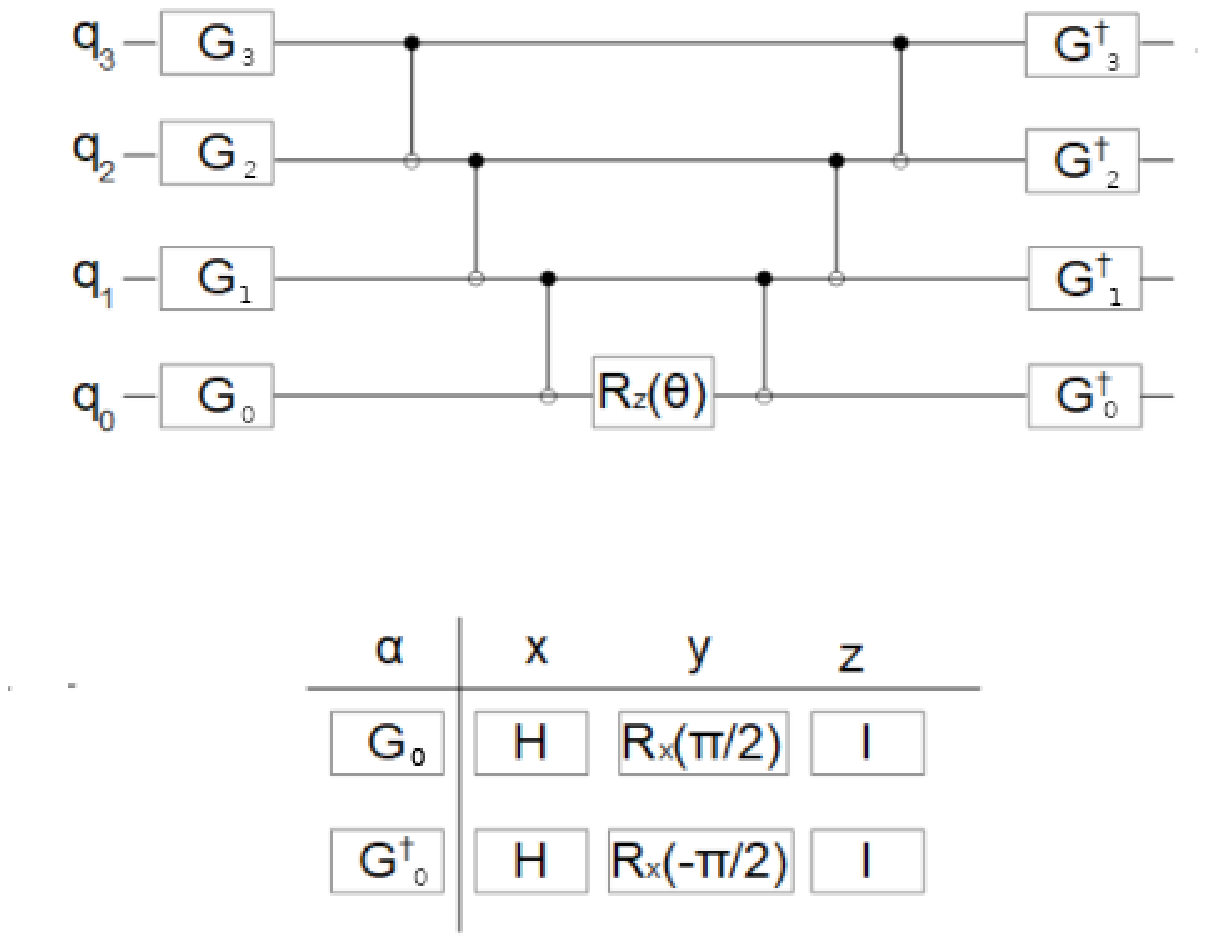}
\caption{Quantum circuit to perform the operation that act on $exp(-i\theta\sigma_0^\alpha\sigma_1^\beta\sigma_2^\gamma\sigma_3^\delta)$.
$\alpha$, $\beta$, $\gamma$, and $\delta$ determine the gate $G_j$ and $G_j^\dagger$ right and left of $q_0$, $q_1$, $q_2$, and $q_3$ corresponding gate for $x$, $y$, $z$ in the list, respectively.}\label{vqeexp}
\end{figure}

The quantum circuit to solve $exp(-i\theta_kP_jt)$ by Hamiltonian is as shown in Fig. \ref{vqeexp}. This quantum circuit is four-qubits system. This quantum circuit acts on multiple term of pauli operater. $q_0-q_3$ indicate first, second, third, and fourth qubit, respectively. $q_0$ and $q_1$ correspond to bonding orbitals and $q_2$ and $q_3$ correspond to anti-bonding orbitals. $R_z(\theta_k)$ corresponds to the variable coefficient in eq.\ref{clus}. $G_j$ and $G^{\dagger}_i$ are eigenvalue-operator and its conjugate operator of pauli operator acted on i-th qubit, respectively. 
For example, $G_0$ become H gate, $G_1$ become $R_x(\pi/2)$ gate and $G_2$ and $G_3$ become I gate if we multiply Hamiltonian by $exp(-i\theta\sigma_0^x\sigma_1^y\sigma_2^z)$. Then, there are not CNOT gate between $q_2$ and $q_3$.
When we multiply Hamiltonian by $exp(-i\theta\sigma_0^x\sigma_2^y\sigma_3^z)$, CNOT gate move from between $q_0$ and $q_1$ to $q_0$ and $q_2$. And, there is no CNOT gate between $q_1$ and $q_2$. $G_1$ disappear, $G_2$ becomes $R_x(\pi/2)$ and $G_3$ becomes $I$ gate. 
By using VQE method we repeat the try and error by acting on the hamiltonian and cluster on quantum computers and change the variable coefficient to optimize in this way. 
The evaluation function of i-th state is represented as,

\begin{eqnarray}
F_i(\bm{\theta}) &=& \langle \Phi_{ini} \mid UHU^\dagger \mid \Phi_{ini} \rangle + E_{i}^{def.}. \label{F}
\end{eqnarray}
Then,
\begin{equation}
U= \Pi_{j,k} exp( i \theta_k P_j t_j)\label{U}.
\end{equation}

$\mid \Phi_{ini}\rangle$  is the ground state $\mid 1000 \rangle$ of the system transformed by Bravyi-Kitaev method. $E_i^{def.}$  and indicate the deflation of i-th state. 
Deflation in eq.\ref{F} become zero when the value of energy is optimized. However, we need to seek the value of energy on excited state. Deflation term is necessary to derive excited states. 
Deflation term is so called Variational Quantum Deflation (VQD) method. VQD method is performed by adding overlap of previous and present states to evaluate wave function. The form is expressed by

\begin{equation}
E(\bm \theta)=\langle\Phi_i\mid(H+A\sum_{j<i}\mid\Phi_j\rangle\langle\Phi_j\mid)\mid\Phi_i\rangle\label{def}.
\end{equation}

Here, A is coefficient to weigh. The overlap of i and j state can be derived by SWAP-test algorithm.
\subsection{Subspace Search VQE}\label{2-2} 
Subspace-Search Variational Quantum Eigensolver (SSVQE) method is advanced algorithm of VQE method that can drive multiple states at once.  As described in introduction, this method automatically generates the orthogonal condition on the output states. The method is optimization method of energy for not each state but summation. The summation of energy is represented by

\begin{eqnarray}
F(\bm \theta)&=&\sum_{j}^{N_d}w_j\langle \Phi_j \mid H \mid \Phi_j \rangle + \sum_j^{N_d}E_j^{def.}\\ \nonumber
&=&\sum_{j}^{N_d}E_j + \sum_j^{N_d}E_j^{def.}\\ \nonumber
When~i&>&j,~E_i>E_j. \\ \label{ss1}
\end{eqnarray}
Here, we calculate for solving minimum energy on the condition of $i>j$, $w_i<w_j$. $w_i$ and $w_j$ are weight constraints.$N_d$ is the number of states that are driven at once. Besides, each initial state of $\mid \Phi_j\rangle$ is different for j. All the cluster should be common on all calculation states. There is some possibility of having low accuracy when excited state cross each other.  Therefore, in this time, we calculate all clusters for each state to keep high accuracy without being trapped by local minimums.
Deflation term is set to be A = 1 for all previous states. Excited states can be derived spontaneously in the order of $i > j$, $E_i > E_j$ when deflation term is zero.

\subsection{Addition to Constraint and Tabu Search}\label{2-3}
$E_i^{const}$ indicate Constraint term of i-th state. Constraint terms can be derived by Lagrange's multiplier method\cite{doi:10.1021/acs.jctc.8b00943} into eq.\ref{F}. The constraint condition of i-th state $E_{i}^{const
}$ is represented by

\begin{equation}
E_{i}^{const}=\sum_{j = 0}^{num~of~const.}\mid \langle \Phi_i \mid ( U_j  - U_{j}^{const})\mid \Phi_i \rangle \mid \label{const}.
\end{equation}

$U_j$ indicates the observable as constraint term and $U_j^{const}$ is targeted value. Here, $U_j$ include $\sum s_i^2$, $\sum s_i^z$ and $N$. Then, the s is spin parameter including magnetic moment that we can observe as constraints and $N$ is the number of electrons. This term is calculated in the same way as for hamiltonian.
We have many local minimums look like Rastregin function when energy lift up to excited state. As the result, electron tend to catch on the local minimums when electron drop in ground state. Therefore, we use Powell method as newtonian method is tend to be trapped on them. 
We add constraint term and Tabu Search terms on excited state. Firstly,  Tabu search term is expressed by 
\begin{eqnarray}
E^{Tabu}&=&\sum_{j}^{num.~of~Tabu.}\exp{(-\mu\langle \Phi_j \mid ( U_j  - U_j^{Tabu.} )\mid \Phi_j \rangle^2)}a \\ \nonumber
&=&\sum_{j}^{num.~of~Tabu.}E_j^{Tabu.}.
\end{eqnarray}

This Tabu Search term is very useful to avoid being trapped by local minimums. Here, $\mu$ is the width and a is the amplitude of tabu Search term in parameter space, respectively. $U_j^{Tabu.}$ indicates the value that must be avoided for $U_j$. If we use the system with degeneration in SSVQE method, the solution that is satisfied with eq.9 emerge according to the number of degeneracy.
To avoid these solutions, The evaluation function is set by, 
\begin{equation}
F_j(\bm \theta)=\langle \Phi_j \mid H \mid \Phi_j \rangle + E^{const.}_j + E^{Tabu}_j.
\end{equation}

This evaluation function is satisfied with $F_i(\bm \theta)>F_j(\bm \theta)$ when i is larger than j. 
We simulate the energy levels of ground state, excited state (triplet, singlet, and doubly) by using VQE and SSQVE with with Constraint and/or Tabu Search terms.
We simulate the initial states on calculation on H$_2$ as $\mid 1000\rangle$, $\mid 0110\rangle$, $\mid 1100\rangle$, and $\mid 0010\rangle$, respectively. And, we solve initial state according to every two levels. Tabu terms are fixed to avoid $\sum s_i^2 = 0.75$, $\sum s_i^z = 10000$,  and $ N = 10000$, respectively.
In the subsection 3.2, we simulate the initial states on calculation on HeH as $\mid 1110 \rangle$, $\mid 1101 \rangle$, $\mid 1011 \rangle$, and $\mid 0111 \rangle$. And, we solve initial state according to every two levels too. Tabu terms are fixed to avoid $N = 10000$.

\subsection{Preparation of calculation}\label{2-4} 
In this time, we use Powell method as classical algorithm. Thenthe number of iterations is limited in 2000 times at a maximum.  Here, $\mu$ is 100 and a is 100 for all evaluation functions. Deflation terms $E^{def.}_i$ and constraint term are as follows. Deflation term of i state $E^{def.}_i$ is expressed by
\begin{eqnarray}
E_{i}^{def.}&=&((af+b(1-f)) \\ \nonumber 
&\times& (\sum_{j<i}(exp(r-0.25r_d)+1)^{-1} \\ \nonumber
&\times & \mid\langle\Phi_j\mid\Phi_i\rangle\mid{^2}\\ \nonumber
&+&(1-(exp(r-0.25r_d)+1)^{-1}) \\ \nonumber
& \times & f(\mid\langle\Phi_j\mid\Phi_i\rangle\mid{^2}))\label{defmisc}.
\end{eqnarray}

a and b, and $f = (exp(\alpha(r - rd)) + 1)^{-1}$ indicate two constants and diatomic bond length r of Fermi-Dirac distribution respectively. Then, $r_d$ is a given diatomic bond length.  a is 1.0 and $\alpha$ is 100.  $f (\mid\langle\Phi_j\mid\Phi_i\rangle\mid^2)$ is inhomogenius function of overlap of i and j states to derive degenerated excited states and is expressed by
\begin{eqnarray}
f(\mid\langle\Phi_j\mid\Phi_i\rangle\mid{^2})&=&(1+2(\sqrt{5}+1))r^4/r_d^4E_p(r)/4\mid\langle\Phi_j\mid\Phi_i\rangle\mid{^4} \\ \nonumber
&+&2(\sqrt{5}+1)r^4/r_d^4E_p(r)/4\mid\langle\Phi_j\mid\Phi_i\rangle\mid{^2}.
\end{eqnarray}

$E_p(r)$ is the value of one lower energy level for given r. All calculations are performed numerically using blueqat SDK\cite{blueqat}.

\section{Numerical Simulation}\label{3}
In this section, we simulate the result of ground and excited states on H$_2$ and HeH.
  We compared constraints with Tabu Search on VQE and SSVQE.  
  We obtain the value of energy state and accuracy of the value. 
  Where, we obtained the relationship between diatomic bond length ($\AA$) vs energy value (Hartree) on each states by simulation. 
  We obtained the relationship between diatomic bond length($\AA$) vs accuracy data. Here, we use $Log_{10}(E-E_{FCI})$ as the indicator of accuracy.
  FCI means Full-CI calculation based on Classic algorithm.

\subsection{The effect of Constrained and Tabu Search term on calculation of H$_2$}\label{4-1}
We show the result of calculation of energy for diatomic bond length on H$_2$ by (1) VQE, (2) constrained VQE, and (3) constrained VQE with tabu Search terms in Fig.3 (A), (B), and (C). 
Moreover, we show the corresponding accuracy of calculation for diatomic bond length in Fig.4 (A), (B), and (C). 
In the case of (1), the error bars of energy calculation increase except for ground state as the distance between hydrogen bond become small as shown in Fig. 3 (A). 
The error bars of energy on calculation of constrained VQE with/without Tabu Search become smaller compared to the calculation on VQE, as shown in Fig.3 (B), and (C).
 However, we cannot recognize the effect on Tabu Search term when the error bar of energy on excited state become larger as shown in Fig.3 (B), and (C).
 We compared the accuracy of each data on VQE method as shown in Fig.4 (A), (B), and (C).  
 The accuracy data of ground state on VQE (Fig.4 (1)) is almost same as that on constrained VQE with/without Tabu Search (Fig.4 (B), and (C)). 
 The accuracy data on ground state are more or less below negative forth power as average as shown in Fig.4 (B). 
 The accuracy data of singlet and doubly excited state on VQE is smallest than other states as shown in Fig.4 (A), (B), and (C).
 The accuracy data of ground state on constrained VQE is smaller than that on conventional VQE method.
 We cannot recognized the effect of Tabu Search (Fig.4 (B), and (C)).

 Next we compared the effect of constrained and Tabu Search in the case of SSVQE.
 We show the result of calculation of energy for diatomic bond length on H$_2$ by (4) SSVQE, (5) constrained SSVQE, and (6) constrained SSVQE with Tabu Search in Fig.3 (D), (E), and (F).
 Moreover, we show the corresponding accuracy of calculation for diatomic bond length in Fig.4 (D), (E), and (F).
 We confirmed the error bar of ground state energy as shown in Fig.3 (D). However, the error bar on constrained SSVQE with/without Tabu search became small compared with that on SSVQE.
 The accuracy data of ground and triplet states on constrained SSVQE with/without Tabu Search is about negative second power on simulated all the range though that on SSVQE is about negative first power.  
 Therefore, we cannot recognize the effect of SSVQE.

From these results, we will select constrained VQE with/without Tabu search for solution of H$_2$. To investigate the effect of Tabu search, we analyze the convergence of energy level as shown in Fig. 5. (a), (b), (c), and (d) shows convergence results in the case of (2), (3), (5), and (6).Tabu search contribute to stabilize the convergence of energy levels. As shown in Fig. 5 (a), (b), singlet and doubly excited states in the case of (2) spike 6 times at most before convergence about in 2100 updates of variables. In contrast, these two states in the case of (3) spike only 5 times at most before convergence in about 1800 updates of variables.

Moreover,  SSVQE method Tabu search contribute to stabilize the convergence more than VQE method with Tabu search. As shown in Fig. 5(c),(d), there are spikes of 11 times in the case of (5) before convergence in 4438 times. In contrast, two states in the case of (6) spike only 3 times before convergence in about 1600 updates of variables. Besides, the improvement of the accuracy of these two states in case of (6) compared to (5) is greater than that of (3) compared to (2). On these VQE and SSVQE, introducing Tabu search contributes to decrease  convergence times on calculation on H$_2$.

\begin{figure*}[h]
\includegraphics[scale=0.58]{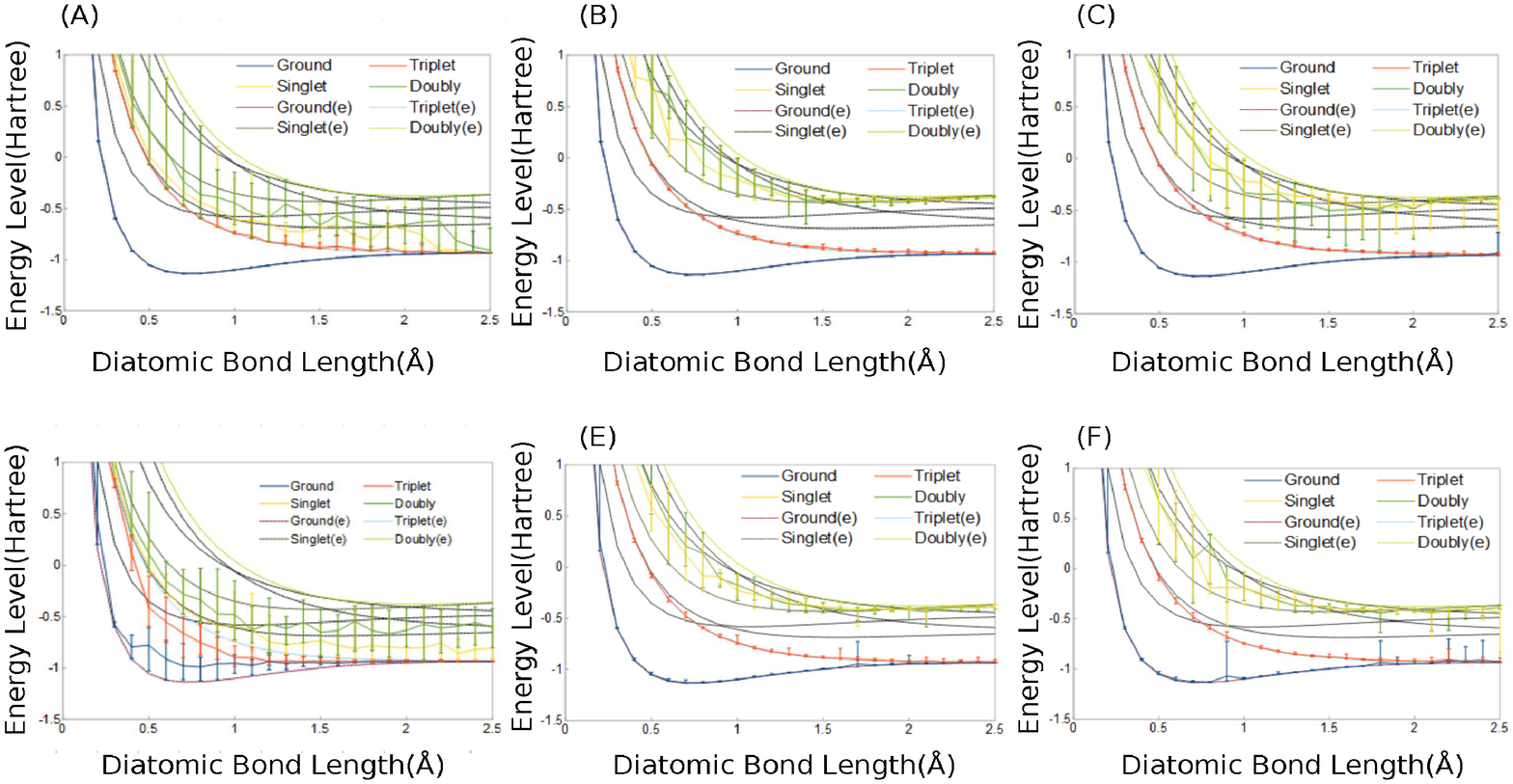}
\newline

\caption{Diatomic bond length($\AA$) of hydrogen molecule v.s.the energy levels (Hartree) of each state calculated by VQE method of the case (1), (2), (3), (4), (5), and (6). Solid line on each state is connecting average points by ten sampling data. (A) Case (1): calculated by VQE method. (B) Case (2): calculated by constrained VQE method. (C) Case (3): calculated by constrained VQE method with Tabu search. (D) Case (4): calculated by SSVQE method. (E) Case (5): calculated by constrained SSVQE method. (F) Case (6): calculated by constrained SSVQE method with Tabu search.
}\label{p0}
\end{figure*}

\begin{figure*}[h]
\includegraphics[scale=0.58]{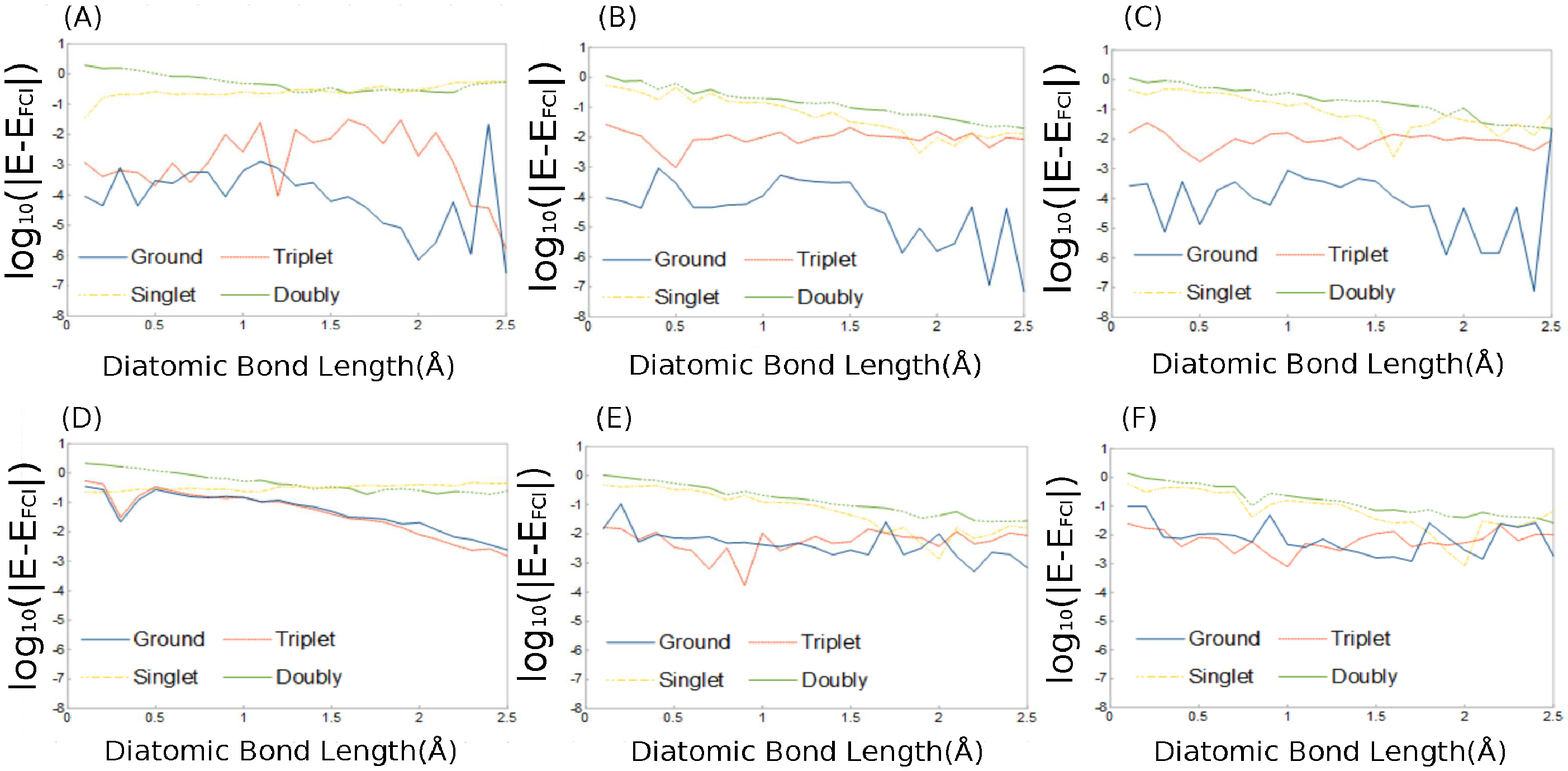}
\newline
\newline
\caption{Diatomic bond length($\AA$) of hydrogen molecule v.s. the accuracy of calculated energy levels ($log_{10}(\mid E - E_{FCI} \mid)$) by VQE method of  the case (1), (2), (3), (4), (5), and (6). Solid line on each state is connecting average points by ten sampling data. (A) Case (1): calculated by VQE method. (B) Case (2): calculated by constrained VQE method. (C) Case (3): calculated by constrained VQE method with Tabu search. (D) Case (4): calculated by SSVQE method. (E) Case (5): calculated by constrained SSVQE method. (F) Case (6): calculated by constrained SSVQE method with Tabu search.
}\label{pa0}
\end{figure*}

\begin{figure*}[h]
\includegraphics[scale=0.55]{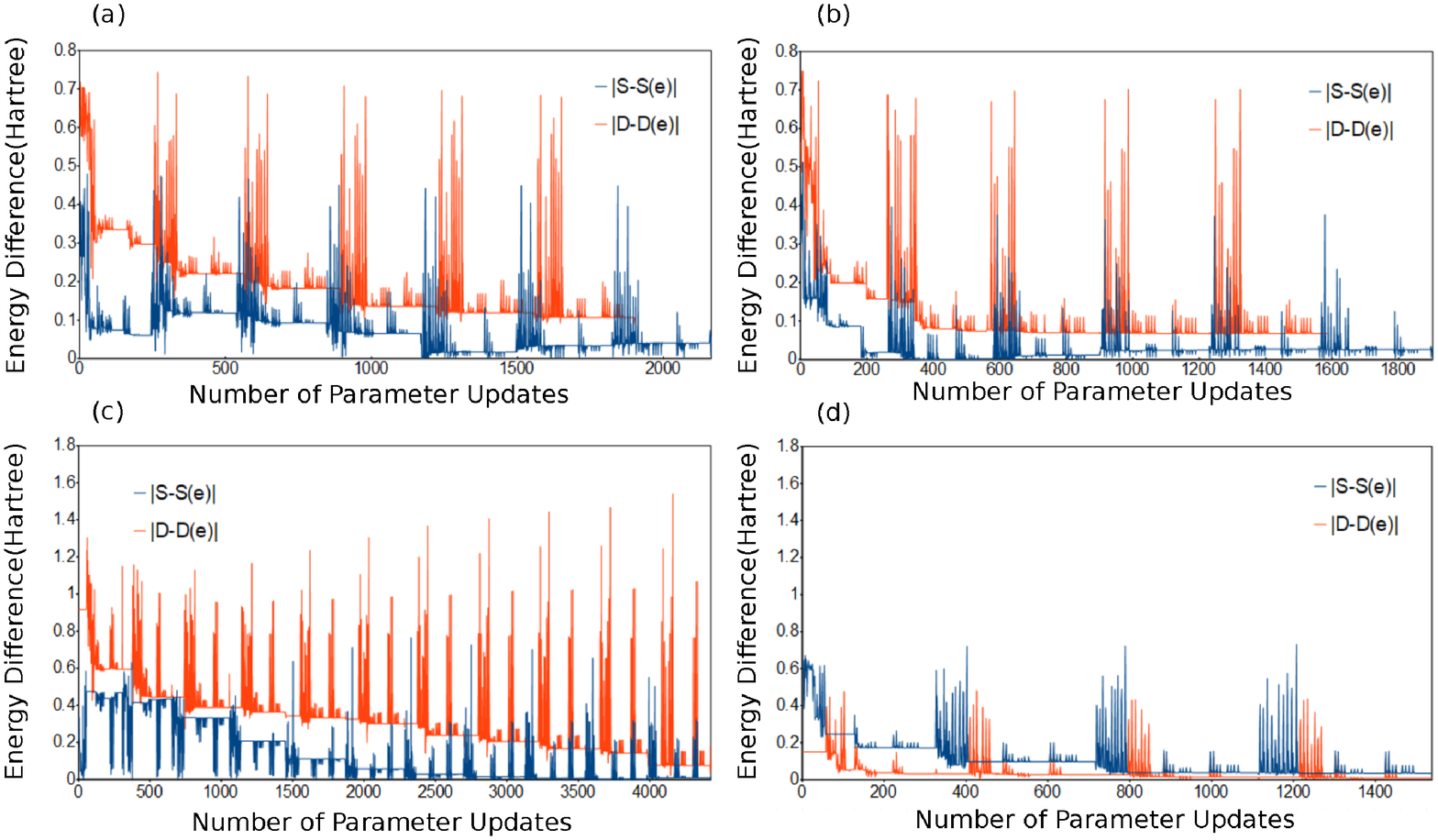}
\newline
\newline
\caption{Number of updates of variables v.s. Energy difference of singlet and doubly excited states of H$_2$ molecules between their grobal minimums, respectively for the case of (2), (3), (5), (6), respectively. (a) number of iteration v.s. Energy difference in the case of (2): constrained VQE. (b) number of iteration v.s. Energy difference in the case of (3): constreained VQE with Tabu search. (c) number of iteration v. s. energy difference in the case of (5): constrained SSVQE. (d) number of iteration v.s. Energy difference in the case of (6): constrained SSVQE with Tabu search.}\label{convH}
\end{figure*}

\subsection{The effect of constrained and Tabu Search term on calculation of HeH}\label{4-2}
 We show the result of calculation of energy for diatomic bond length on HeH by (7) VQE, (8) constrained VQE, and (9) constrained VQE with Tabu Search terms in Fig.6 (A), (B), and (C).
 Moreover, we show the corresponding accuracy of calculation for diatomic bond length in Fig.7 (A), (B), and (C).
 In the case of (7), (8) and (9), errors bar on excited states became huge large though we could not find errors on ground states 1. 
 The behavior (-4 as log) of ground state 1 is almost same as that on H$_2$ on VQE method. On the other excited states, the accuracy is about 0 to -2.  

 We show the result of calculation of energy for diatomic bond length on HeH by (10) SSVQE, (11) constrained SSVQE and (12) constrained SSVQE with tabu Search terms in Fig.6 (D), (E), and (F).
 Moreover, we show the corresponding accuracy of calculation for diatomic bond length in Fig.7 (D), (E), and (F).
 We cannot find error bar on each energy states when we use the method of SSVQE.
 When diatomic bond length is beyond 0.5 $\AA$, accuracy became below negative third power on ground and excited 2 states.
 Moreover, the accuracy data on constrained SSVQE with/without Tabu Search became negative third power to negative forth power on all the range beyond 0.5 $\AA$. 
 Here, we could not recognize the difference with/without Tabu Search only for results.

As described above, energy error obtained by SSVQE is smaller than that by VQE for HeH. To investigate convergence condition, we analyze the convergence of energy of all the methods. Fig. 8 shows energy difference vs iteration in the case of (7), (8), (9), (10), (11) and (12). Fig. 8 (a) denotes VQE method, (b) constrained VQE method, and (c) constrained VQE method with Tabu Search, for HeH, respectively. Fig. (d) denotes SSVQE, (e) constrained SSVQE method, and (f) constrained SSVQE with Tabu Search, for HeH, respectively.  
As shown in Fig. 8 (a), (b), (c), doublet excited states in the case of (7) and (8) spike 3 times before convergence in 1400 updates of variables. In contrast, these states in the case of (9) spike only 2 times before convergence in about 900 updates of variables. Besides, the energy levels of these two states approach global minimum, much faster than those of case (8). Tabu search term contribute to stabilize the convergence on VQE and SSVQE methods for analysis of energy level of HeH.  
As shown in Fig. 8 (d), (e), (f), especially those energy levels in the case of (10) and (11) spike 3 times before convergence in about 1500 updates of variables. In contrast, those two states in the case of (12) spike only 2 times before convergence in about 1000 updates of variables. Besides, one of these state (E2) approach global minimum twice faster than that in the case of (11). Intensity of spikes is weaker than that in the case of (11) too. On these VQE and SSVQE, introducing Tabu search contributes to decrease  convergence times on calculation on HeH too.

\begin{figure*}[h]
\includegraphics[scale=0.55]{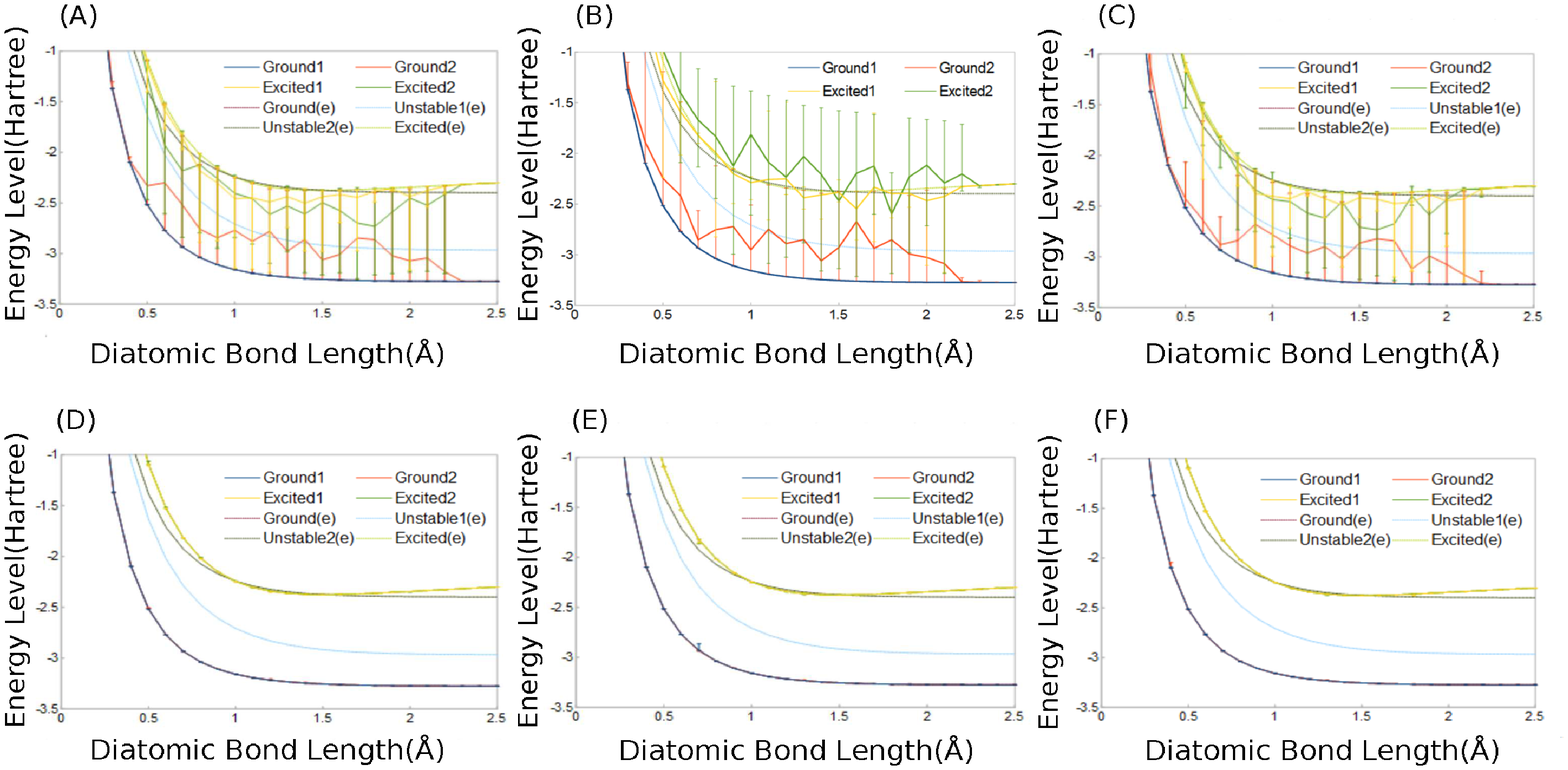}
\newline
\caption{Diatomic bond length($\AA$) of heliuym hydrite v.s.the energy levels (Hartree) of each state calculated by VQE method of the case (7), (8), (9), (10), (11), and (12). Solid line on each state is connecting average points by ten sampling data. (A) Case (7): calculated by VQE method. (B) Case (8): calculated by constrained VQE method. (C) Case (9): calculated by constrained VQE method with Tabu search. (D) Case (10): calculated by SSVQE method. (E) Case (11): calculated by constrained SSVQE method. (F) Case (12): calculated by constrained SSVQE method with Tabu search.
}\label{pHeH0}
\end{figure*}

\begin{figure*}[h]
\includegraphics[scale=0.55]{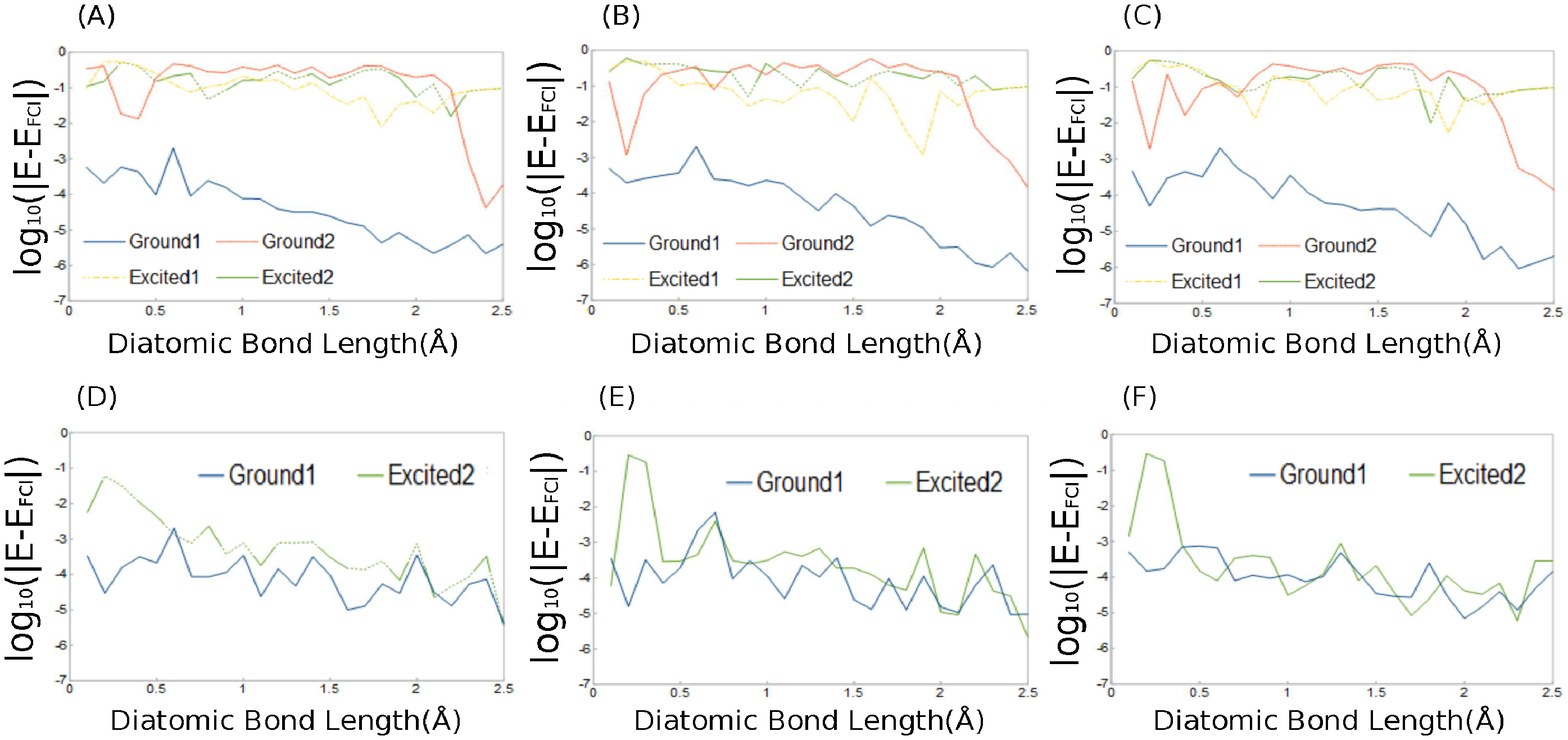}
\newline
\newline
\newline
\caption{Diatomic bond length($\AA$) of heliuym hydrite v.s. the accuracy of calculated energy levels ($log_{10}(\mid E - E_{FCI} \mid)$) by VQE method of  the case (7), (8), (9), (10), (11), and (12). Solid line on each state is connecting average points by ten sampling data. (A) Case (7): calculated by VQE method. (B) Case (8): calculated by constrained VQE method. (C) Case (9): calculated by constrained VQE method with Tabu search. (D) Case (10): calculated by SSVQE method. (E) Case (11): calculated by constrained SSVQE method. (F) Case (12): calculated by constrained SSVQE method with Tabu search.
}\label{paHeH0}
\end{figure*}

\begin{figure*}[h]
\includegraphics[scale=0.55]{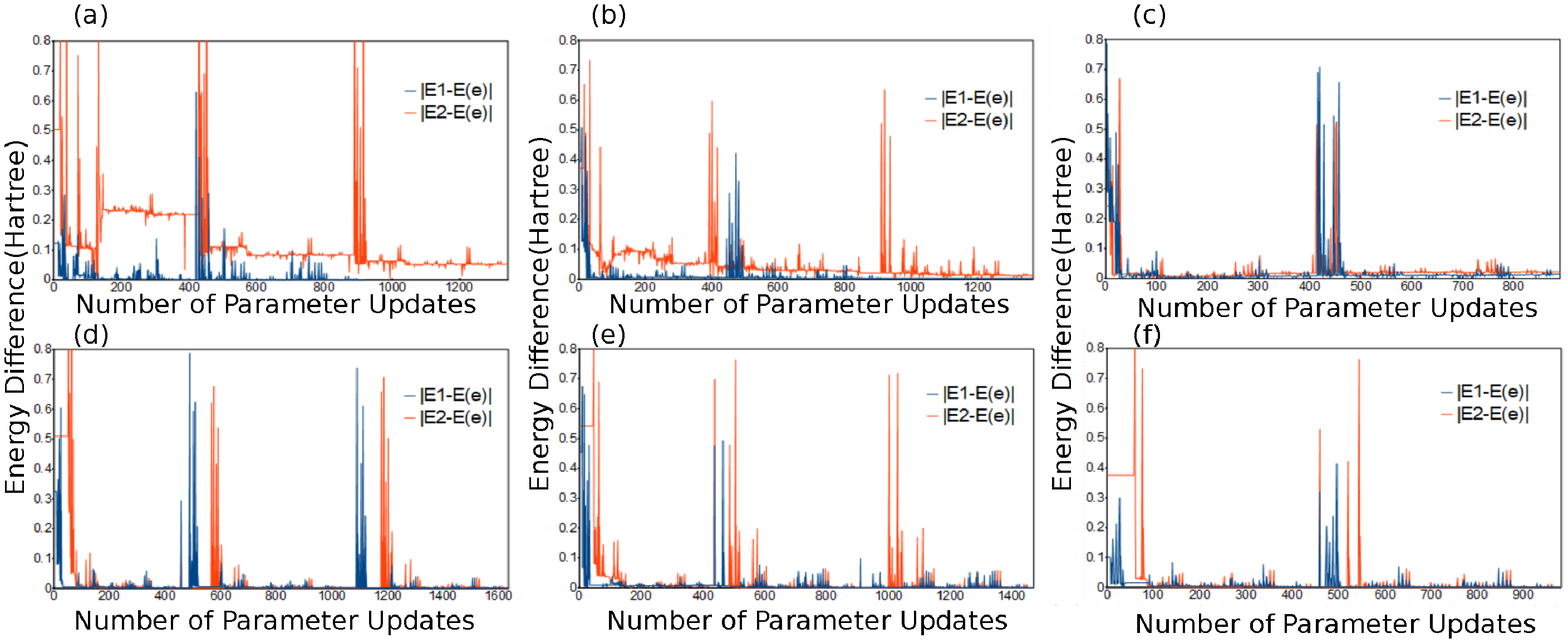}
\newline
\newline
\caption{Number of updates of  parameters v.s. Energy difference of doublet excited states of HeH molecules between their grobal minimums, respectively for the case of (7), (8), (9), (10), (11), and (12), respectively. (a) number of iteration v.s. Energy difference in the case of (7): VQE method. (b) number of iteration v.s. Energy difference in the case of (8): constrained VQE. (c) number of iteration v.s. Energy difference in the case of (9): constreained VQE with Tabu search. (d) number of iteration v.s. Energy difference in the case of (10): SSVQE method. (e) number of iteration v. s. energy difference in the case of (11): constrained SSVQE. (f) number of iteration v.s. Energy difference in the case of (12): constrained SSVQE with Tabu search.}\label{convHeH}
\end{figure*}

\section{Discussion}\label{4}
Our purpose is to obtain fine accuracy trace relationship between bonding distance and energy on excited energy at the distance not to drop in local minimum. The reason is because we need accurate physical and chemical constant for chemical reaction. 

VQE is developed for obtaining ground state energy. On the other hand, SSVQE is developed for obtaining excited state energy. The error of ground state energy is fewer than that of excited energy on VQE on H$_2$. Moreover, the error of excited energy is fewer than that of ground state on SSVQE on H$_2$. The error on HeH exhibits a similar tendency on VQE. However, the error on SSVQE on HeH is different from VQE on H$_2$ and it is very small.
The ground and excited states are both doublet on HeH. Hence, two degenerated states are derived at once.
Therefore, we think that it is important to fill degenarated orbitals or create electron pair.

H$_2$ molecules is stable as H atom consist of one proton and one electron and H$_2$ molecules has 2s orbital that filled with two electrons. As the results, analysis of energy state is enough to simulate on VQE. HeH is material produced by nature and stable as positive ion.  Excited energy system is thought to be little bit unstable. To avoid the unstable states we need to give restraint that is called constrained term. As the results, constrained term is useful for analysis of ion. To obtain fine accuracy, we had better use SSVQE method. 

The constraint terms was beneficial for enhancing  the accuracy as well as decreasing errors. 
For the reasons mentioned above, Tabu search term was useful by VQE methods on H$_2$ and SSVQE method on HeH so as to decrease convergence times. The reasson is because Tabu search is developed as  metaheurestic search algorithms, thus, Tabu search terms cut off the paths destinated in local minimums in order to updates parameter sets pass the shortest way to grobal minimum.

 We are considering the charge balance on optimization depends on deviation of whole molecules. We think that positive charge have an effect on the balance of charge on HeH.

\section{Concluding Remarks}\label{5}
We investigated the effect of constraint and Tabu Search term on VQE and SSVQE for ground and excited states. As a result, VQE method is adequate for the calculation on H$_2$ and SSVQE is adequate for the calculation on HeH, respectively. Constraint and Tabu search terms contribute to enhance the accuracy of energy levelsaccording to each bond length and to decrease convergence times.  According to the nature of molecules, we had better to use VQE or SSVQE with constraint and Tabu search terms.Therefore, we suppose to select adequate method (VQE or SSVQE) considering electron charge condition on outer orbital.

Next subject is to improve the accuracies on calculation when objective molecules have different orbital set. In order to achieve it, we had better modify initial states and append extra terms though their trials are formidable. Simplifying cluster and applying other optimizer that can search the grobal minimum avoiding local them\cite{2006}\cite{2020arXiv200704424C}\cite{Kento2013} are also. As a next step, it is also worth to optimize energy levels of states by using Tabu search on larger molecules. We will search suitable quantum algorithm such as VQE, SSVQE, constraint and Tabu search term and so on when we solve energy state on large molecules. We will select suitable method that depend on the nature of molecules such as electron pair, balance of charge.
\newpage
\bibliographystyle{apsrev4-2}
\bibliography{temp4}
\end{document}